\documentclass[aps,pra,twocolumn,superscriptaddress,notitlepage,floatfix]{revtex4-2}

\usepackage[T1]{fontenc}
\usepackage{graphicx}
\usepackage{verbatim}
\usepackage[separate-uncertainty=true]{siunitx}
\usepackage{footnote}

\usepackage{amsmath}
\usepackage{amssymb}
\usepackage{comment}
\usepackage{soul}

\usepackage{hyperref}
\hypersetup{colorlinks, citecolor=blue, linkcolor=blue, urlcolor=blue}

\newcommand{\ket}[1]{\lvert#1\rangle}

\newcommand{\comm}[1]{}

\DeclareSIUnit{\Isat}{I_{sat}}

\begin{document}
\title{Spatial calibration of high-density absorption imaging}

\author{T.\,Vibel}
\author{M.\,B.\, Christensen}
\author{M.\,A.\,Kristensen}
\author{J.\,J.\,Thuesen}
\author{L.\,N.\,Stokholm}
\affiliation{Center for Complex Quantum Systems, Department of Physics and Astronomy, Aarhus University, Ny Munkegade 120, DK-8000 Aarhus C, Denmark.}
\author{C.\,A.\,Weidner}
\affiliation{Quantum Engineering Technology Laboratories, H. H. Wills Physics
Laboratory and Department of Electrical and Electronic Engineering,
University of Bristol, Bristol BS8 1FD, United Kingdom}
\author{J.\,J.\,Arlt}
\affiliation{Center for Complex Quantum Systems, Department of Physics and Astronomy, Aarhus University, Ny Munkegade 120, DK-8000 Aarhus C, Denmark.}

\date{\today}

\begin{abstract}
The accurate determination of atom numbers is an ubiquitous problem in the field of ultracold atoms. For modest atom numbers, absolute calibration techniques are available, however, for large numbers and high densities, the available techniques neglect many-body scattering processes. Here, a spatial calibration technique for time-of-flight absorption images of ultracold atomic clouds is presented. The calibration is obtained from radially averaged absorption images and we provide a practical guide to the calibration process. It is shown that the calibration coefficient scales linearly with optical density and depends on the absorbed photon number for the experimental conditions explored here. This allows for the direct inclusion of a spatially dependent calibration in the image analysis. For typical ultracold atom clouds the spatial calibration technique leads to corrections in the detected atom number up to $\approx\! \SI{12}{\percent}$ and temperature up to $\approx \!\SI{14}{\percent}$ in comparison to previous calibration techniques. The technique presented here addresses a major difficulty in absorption imaging of ultracold atomic clouds and prompts further theoretical work to understand the scattering processes in ultracold dense clouds of atoms for accurate atom number calibration.
\end{abstract}

\maketitle

\section{Introduction}
\label{sec:Introduction}

The accurate measurement of the atom numbers and temperatures of clouds of ultracold atoms is rife with complications, both in the case of dilute mesoscopic samples and for samples at high densities. In almost all cases, detection relies on optical methods which can be divided broadly, in resonant and dispersive techniques. 

Resonant techniques rely on the detection of scattered photons, and the most common are absorption and fluorescence imaging, where the atom number is inferred from the number of resonant photons scattered by an atomic sample. Fluorescence methods, which rely on direct capture of these scattered photons, have been used ubiquitously in atomic physics. Absorption imaging is a workhorse in the field of ultracold atomic physics, including the investigation of Bose-Einstein condensation~\cite{Ketterle1999}, but its accurate calibration remains difficult~\cite{Reinaudi2007, Veyron2022}.  Other variations on absorption imaging exist, including resonant frequency modulation imaging~\cite{Bjorklund1983}, which has been shown to produce images with high signal-to-noise ratio (SNR)~\cite{Robins2016}. Absorption imaging can also be implemented in a \textit{dark-ground} configuration using spatial filtering in the Fourier plane of the imaging setup to remove the unscattered probe beam and thus improve SNR~\cite{Pappa2011,Reinhard2014}. There also exist minimally-destructive, but still resonant, methods to perform absorption imaging of atomic clouds, including partial-transfer absorption imaging~\cite{Campbell2012}. 

Dispersive techniques, on the other hand, detect a phase shift imprinted on the off-resonant probe light by the spatially varying refractive index of the atomic sample. Since these techniques do not rely on scattering of photons they can be made minimally destructive and are suitable for repeated imaging of ultracold samples. Techniques that have been applied to cold and ultracold atoms include phase-contrast imaging~\cite{Ketterle1996, Hulet1997, Ketterle1997, vdS2010}, off-resonant frequency modulation spectroscopy~\cite{Aspect1999}, spatial heterodyne imaging~\cite{Walker2001}, diffraction contrast (or defocus) imaging~\cite{Scholten2003, Scholten2005, Kuhn2016, Spielman2021}, and Faraday imaging~\cite{Gajdacz2013, Eliasson2019}.

In absorption imaging, an atomic sample is probed by resonant light and the atomic density is inferred based on the photons scattered out of the probing beam. For a cloud of two-level atoms in a monochromatic light field the attenuation can be obtained from the steady-state solution of the optical Bloch equations~\cite{Straten2016}
\begin{equation}\label{eq:ScatteringRate}
	\frac{dI}{dz} = -\frac{\sigma_0 n(z)}{1 + I/I_\mathrm{sat} + \left(\tfrac{2\delta}{\Gamma}\right)^2} I,
\end{equation}
where $n(z)$ is the atomic density, $I_\mathrm{sat}$ is the saturation intensity, $\delta$ is the laser detuning from the atomic transition and $\Gamma$ is the line width of the transition.

The resonant scattering cross-section is related to the saturation intensity by $\sigma_0 = \hbar\omega_0\Gamma/2 I_\mathrm{sat}$ where $\omega_0$ is the resonance frequency. After integration along the $z$-direction, the optical density ($od$) and the column density $\tilde{n}$ are given by~\cite{Reinaudi2007} 
\begin{equation}
	od = \tilde{n}\sigma_0 = \left(1+\frac{4\delta^2}{\Gamma^2}\right)\ln\left(\frac{I_0}{I_\mathrm{p}}\right) + \frac{I_0-I_\mathrm{p}}{I_\mathrm{sat}},
	\label{eq:LambertBeerSolution}
\end{equation}
where $\tilde{n} = \int\mathrm{d}z\, n(z)$. 
Extracting the column density requires a probe and a reference image, both corrected for dark counts in the absence of probing light. The probe image $I_\mathrm{p}(x,y) $ contains the shadow cast by the atoms while the reference image $I_0(x,y)$ contains an image of the probe beam in the absence of the atoms. 

Accurate detection relies on correct determination of the saturation intensity $I_\mathrm{sat}$, which will generally differ from the result of the two-level atom due to the complex multi-level structure of the internal states of real atoms and imperfect polarization states of the probe beam. Additionally, the probing light imparts momentum to the atoms, which can lead to an effective detuning for typical probing durations. At higher densities, additional effects such as re-absorption processes and coherent many-body scattering come into play~\cite{Veyron2022, Veyron2022b}. One must also correct for the effect of an imperfect probe pulse shape~\cite{Horikoshi2017}. All of these preclude an accurate determination of the number of atoms in a dense, cold cloud of atoms.

At low density and atom number $\lesssim \num{1e4}$, however, some calibration techniques have been developed. Single-atom resolved detection allows for the observation of step-like features in the atomic signal, which enables direct counting of atom numbers~\cite{Streed2012,Hume2013,Hueper2019}. This approach is most often applied to fluorescence imaging, as it provides a better SNR at low atom numbers. At a few hundred atoms, spatially non-resolved single-atom counting becomes challenging. However, the linear relationship between the quantum projection noise and the mean atom number of coherent spin states has been used to ensure accuracy in atom counting~\cite{Muessel2013}.

At higher atom numbers, it has been shown that the effective saturation intensity can be inferred from the momentum acquired by the atoms during imaging~\cite{Hueck2017}  if an auxiliary imaging axis is available. One can also measure the number of photons absorbed by atoms during optical pumping~\cite{Yu2001, Yu2007} to obtain an estimate of the atom number. However, the most generally employed method for calibration takes advantage of the fact that the measured atomic density should not depend on the intensity of the probe beam. By imaging samples at a given density with different probe intensities, the saturation intensity $ I_\mathrm{sat} $ can be adjusted to provide accurate atom numbers~\cite{Reinaudi2007}. 

In this work, we apply the calibration technique presented in Ref.~\cite{Reinaudi2007} to ultracold atomic clouds, but our method varies the calibration as a function of optical density, allowing one to probe density-dependent effects. Our experimental apparatus is in a unique position to perform these measurements since it can produce thermal samples actively stabilized in atom number at the shot noise level~\cite{Gajdacz2016}. This is achieved through a combination of non-destructive Faraday imaging, online image analysis and active control of the atom number. Based on these measurements, we observe a dependence of the saturation intensity $ I_\mathrm{sat} $ on optical density and the number of photons absorbed per atom. A simple recipe is presented to incorporate this calibration method into the evaluation of absorption images. Under typical experimental conditions, it leads to corrections in the detected atom number of up to $\approx\!\SI{12}{\percent}$ and in the temperature of up to approximately $\approx\!\SI{14}{\percent}$. The lack of active stabilization does not preclude the calibration we describe here. Rather, the atom number fluctuations common to non-stabilized experiments will simply lead to larger uncertainties in the calibrated values. Note that the experiments presented here are related to recent \emph{in-situ} measurements~\cite{Veyron2022}; however, Ref.~\cite{Veyron2022} does not provide an analysis for the commonly used time-of-flight imaging technique.

\section{Experimental method \label{sec:expt}}

The calibration measurements were performed in an experimental apparatus designed to produce and investigate Bose-Einstein condensates, which was previously described in detail~\cite{Kristensen2017}; relevant aspects are summarized here. Initially, $\approx\!10^9 $ $^{87}\mathrm{Rb}$ atoms are trapped and cooled in a magneto-optical trap from a background vapor. Subsequently, the atoms are transferred to a magnetic quadrupole trap and mechanically moved to a second chamber with lower pressure and longer lifetime of the atomic clouds. There, the atoms are confined in an elongated quadrupole Ioffe-configuration trap~\cite{Esslinger1998} with trapping frequencies $\omega_r/2\pi = \SI{247.5(2)}{\hertz}$ and $\omega_z/2\pi = \SI{17.26(1)}{\hertz}$ and further cooled by forced radio-frequency~(RF) evaporative cooling. 

The evaporative cooling sequence consists of a series of frequency sweeps. When the cloud contains $N \sim 4\times10^6$ atoms at a temperature of $\sim\SI{14}{\micro\kelvin}$, the sequence is briefly interrupted and the cloud is probed \emph{in situ} by minimally-destructive Faraday imaging~\cite{Gajdacz2013}. The dispersive Faraday interaction leads to a rotation of the polarization proportional to the atomic column density $\tilde{n}$ and allows for probing of the atom number with sub-shot noise precision~\cite{Kristensen2017}. These Faraday images are analyzed in real-time on a field-programmable gate array (FPGA), which also calculates the required fraction of atoms that must be removed to reach the desired target atom number. To induce the required losses, the FPGA controls an RF-synthesizer and applies a loss pulse with a duration adjusted to discard the appropriate number of excess atoms. The frequency of the RF radiation is chosen to remove atoms at the mean energy of the sample and thus minimize the change in temperature. Thus, atom number stability at the shot noise level can be achieved~\cite{Gajdacz2016}. To validate successful stabilization, a second set of $20$ Faraday images is acquired. Subsequently, the cloud is held in-trap for $\SI{300}{\milli\second}$ to allow for thermal equilibrium to be established. The resulting atomic clouds contain $3.3\times10^6$ atoms at a temperature of \SI{2.0}{\micro\kelvin}, which corresponds to a factor of three above the critical temperature to obtain Bose-Einstein condensation.

\subsection{Absorption imaging}
\label{sec:Absorption}

The spatially-dependent calibration technique presented here is applied to absorption images, which are acquired as follows. To initiate the time-of-flight (ToF) imaging procedure, the magnetic trapping potential is extinguished and the atomic cloud expands freely under the influence of gravity. During the time of flight, a homogeneous magnetic field pointing parallel to the imaging beam direction is applied, which enables imaging on the closed-cycle $ \sigma_+ $-transition from the $\ket{F=2, m_F = 2}$ state to the $\ket{F'=3, m_{F'} = 3}$ state. The clouds are imaged after \SI{12}{\milli\second} of time of flight, which results in a maximal optical density $ \tilde{n}_\mathrm{max}\sigma_0 \approx 4.5$.

The imaging setup consists of two lens pairs with a total magnification of $M = \num{4.3}$. The first lens pair forms an intermediate image which allows us to restrict the imaging beam path from all sides using a four-blade knife edge mask. This ensures that the imaging light exclusively falls on the CCD chip and thus reduces scattered imaging light in particular from other parts of the CCD camera.

The imaging pulses are generated with an acoustic-optic modulator (AOM) in a double-pass configuration. The driving RF power is chosen to saturate the AOM, which minimizes the thermal drift of the deflected beam during the first few microseconds after turning it on. This allows us to produce reliable square pulses with durations as short as \SI{3}{\micro\second}. 

To reduce the effect of vibrations of the optical elements, the separation between the probe and reference images is reduced to a minimum. This is achieved by applying an optical pumping pulse for a duration of a $\SI{200}{\micro\second}$ on the $\ket{F=2}$ to $\ket{F'=2}$ transition, which transfers the atoms to the dark $\ket{F=1}$ state.

In total, four images are recorded during absorption imaging. The first two contain the absorption signal of the atoms $I_\mathrm{p}$ and the reference image $I_0$, and are separated by \SI{340}{\micro\second}. Importantly, we compare the average intensity in the probe and the reference image in a region that is free from atoms and apply a correction factor to the reference image that balances these intensities. 
This accounts for small differences in the number of photons in the probe and reference pulse. 
Furthermore, two dark images are recorded three seconds later, where the imaging light is blocked. These images are used to correct for stray light reaching the camera, which may vary over time as e.g., room lights are switched on and off.

\subsection{Calibration technique}

To calibrate absorption imaging, an effective saturation intensity $I_\mathrm{sat} \rightarrow I_\mathrm{sat}^\mathrm{eff} = \alpha I_\mathrm{sat}$ was introduced by Reinaudi et al.~\cite{Reinaudi2007}, which can mitigate some of the effects outlined in Sec.~\ref{sec:Introduction}. This allows a detuning of the imaging light to be absorbed in the $\alpha$ coefficient as $\alpha^* = \alpha\left(1+\ 4\delta^2/\Gamma^2\right)$. Inserting this into Eq.~\eqref{eq:LambertBeerSolution} and using the implicit dependence of the cross-section on $I_\mathrm{sat}$, the expression for the optical density becomes
\begin{equation}
\label{eq:LambertBeerSolutionWithAlpha}
od = \alpha^*\ln\left(\frac{\tilde{I}_0}{\tilde{I}_\mathrm{p}}\right) + \tilde{I}_0-\tilde{I}_\mathrm{p}
\end{equation}
where the normalized quantities are $\tilde{I}_{j} = I_{ j}/I_\mathrm{sat}$, and from here on, the index $j$ refers to the probe (p) or reference (0) image. At low intensity, $ \tilde{I}_0\ll 1 $, the optical density is primarily given by the logarithmic term $ \tilde{n}\sigma_0 \approx \alpha^*\ln\left(\tilde{I}_0/\tilde{I}_\mathrm{p}\right)$, while at large incident intensities $ \tilde{I}_\mathrm{0} \gtrsim 1 $, the linear term becomes relevant or even dominant. Since the latter term depends directly on the imaging light intensity, a calibration of the imaging camera is required. 

The number of photons in the object plane of the atom cloud corresponding to the $\ell$-th pixel is given by
\begin{equation}
    \mathcal{N}_{j}^{(\ell)} = \frac{\mathcal{N}_{\mathrm{c},j}^{(\ell)}}{\mathcal{T}\eta g}
\end{equation}
where $\mathcal{N}_{\mathrm{c},j}$ is the number of photons counted by the camera, $\mathcal{T}$ is the transmission through the optical elements in the imaging system, $\eta$ is the quantum efficiency of the camera chip and $g$ is the camera gain. Here, we assume that these images are dark-image corrected. The average normalized intensity in a given pixel is given by
\begin{equation} 
\label{eq:Intensity}
    \tilde{I}_j^{(\ell)} = \frac{I_j^{(\ell)}}{I_\mathrm{sat}} = \frac{\mathcal{N}_{j}^{(\ell)} \hbar \omega}{I_\mathrm{sat} \tau A}	= C \cdot \frac{\mathcal{N}_{\mathrm{c},j}^{(\ell)}}{\tau}
\end{equation}
where $\tau$ is the imaging pulse duration, $A$ is the pixel area in the object plane and $C = \hbar \omega / (\mathcal{T} \eta g I_\mathrm{sat} A)$ is the camera calibration that converts camera counts to imaging intensity.

In addition, this analysis provides the average number of absorbed photons per atom
\begin{align}
   \overline{\mathcal{N}}_\mathrm{abs} = \frac{1}{N}
    \sum_\ell^{\mathrm{ROI}} \mathcal{N}_{0}^{(\ell)} - \mathcal{N}_{p}^{(\ell)}
\end{align}
where $N$ is the total atom number $N = A/\sigma_0 \sum_\ell^\mathrm{ROI} od^{(\ell)}$. The sum is over all pixels $\ell$ in the region of interest (ROI) of the image. 

To infer $\alpha^*$, one takes advantage of the low- and high-intensity limits of Eq.~\eqref{eq:LambertBeerSolutionWithAlpha}. In practice, the intensity of the imaging light $\tilde{I}_0$ is varied from shot to shot while keeping the density of the imaged samples constant. Thus, one obtains data where the balance between the linear and logarithmic terms in Eq.~\eqref{eq:LambertBeerSolutionWithAlpha} varies. Since the optical density should not depend on intensity, the optimal value of $\alpha^*$ is the one that leads to the smallest variance of the evaluated optical density. However, the optimal $\alpha^*$ depends on density and on the number of absorbed photons, as shown in the following.

\section{Spatially dependent $\alpha^*$-calibration}
\label{sec:Spatialcalib}

\begin{figure}[tbp]
	\centering
	\includegraphics[width=0.8\linewidth]{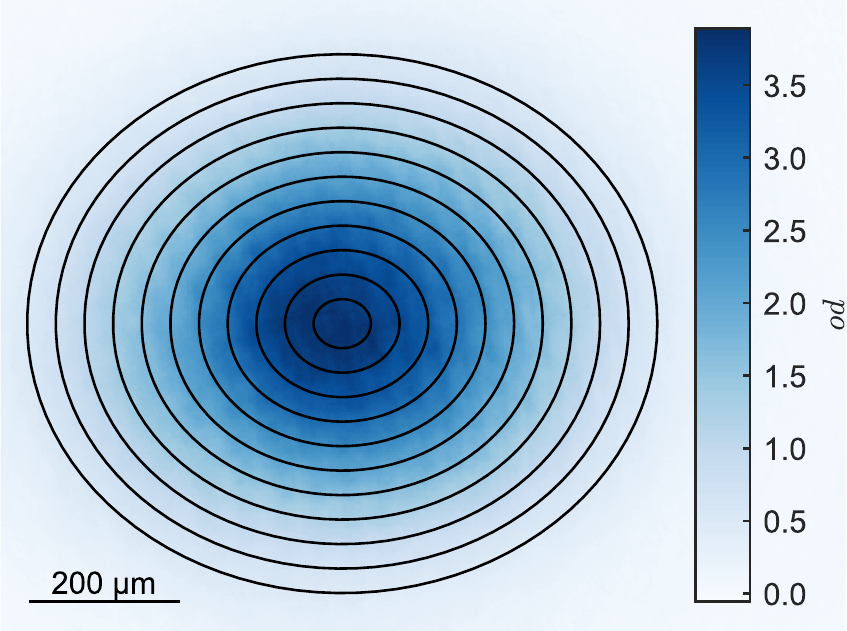}
        \caption{Optical density of an atomic cloud recorded for standard imaging conditions using $\alpha^*=1$. Lines indicate borders of elliptical regions in which separate values of $\alpha^*$ are extracted.
	}
	\label{fig:1ROI}
\end{figure}

To obtain a spatially corrected $ \alpha^* $-coefficient, the two dark-count-corrected images, $ I_\mathrm{p} $ and $ I_0$, are azimuthally averaged. The aspect ratio of the cloud is taken into account by averaging elliptical ROIs with a width of 10 pixels, where the major and minor axes of the ellipses are proportional to the cloud widths in both directions. These cloud widths are determined by a Gaussian fit to the cloud $od$ given by~Eq.~\eqref{eq:LambertBeerSolutionWithAlpha} with $ \alpha^* = 1 $. 
Figure~\ref{fig:1ROI} shows the optical density of a single, typical experimental image of a thermal atom cloud with its Gaussian density profile and these elliptical ROIs. The widths of the elliptical ROIs were chosen as a compromise between the desired spatial resolution and the signal-to-noise ratio of the resulting calibration coefficients.

Images like the one shown in Fig.~\ref{fig:1ROI} are acquired for a range of imaging intensities at a fixed average number of absorbed photons $\overline{\mathcal{N}}_\mathrm{abs}$. This minimizes the influence of, e.g., the induced Doppler shift from scattered photons. When performing a single calibration, $\overline{\mathcal{N}}_\mathrm{abs}$ should match that used for actual experiments. Consequently, the probe exposure time has to be adjusted to account for the imaging intensity. The maximum available intensity was $7.9~I_\mathrm{sat}$, resulting in a minimum pulse duration of $2.2~\mathrm{\mu s}$. On the other hand, the measurements were performed with pulse durations below $32.3~\mathrm{\mu s}$ to constrain the vertical movement of the atoms during probe exposure to the size of one pixel. This results in a minimum imaging intensity of $0.13~I_\mathrm{sat}$. For each imaging intensity, 2 to 3 experimental runs are performed.

To obtain the $\alpha^*$-coefficient, Eq.~\eqref{eq:LambertBeerSolutionWithAlpha} can be recast into a linear equation
\begin{equation} 
\langle \tilde{I}_0-\tilde{I}_\mathrm{p} \rangle_\mathrm{k} =  \langle od \rangle _\mathrm{k} -  \alpha^*_\mathrm{k} \bigg\langle \ln\left(\frac{\tilde{I}_0}{\tilde{I}_\mathrm{p}}\right) \bigg\rangle_\mathrm{k},
\label{eq:LambertBeerLinear}
\end{equation}
where the $\langle \rangle_\mathrm{k}$ indicates averages over the k-th elliptical ROI. Since these averages can be determined directly from the image data, a linear fit can be used to determine both the $\alpha^*_\mathrm{k}$-coefficient and the optical density $od_\mathrm{k}$ in each ROI.

\begin{figure}[tbp]
	\centering
	\includegraphics[width=1\linewidth] {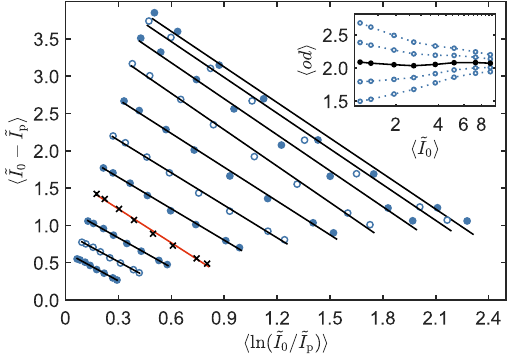}
        \caption{Determination of the $\alpha^*_\mathrm{k}$-coefficient and optical density $od_\mathrm{k}$. The data points in each region of interest $\mathrm{k}$ (alternating blue dots and circles) show a linear dependence. The black lines are fits according to Eq.~\eqref{eq:LambertBeerLinear} where the slope corresponds to the value of $\alpha^*$. The data points indicated by crosses and the corresponding linear fit (red line) represent a global determination of $\alpha^*$ averaging over the entire image. For this calibration, the average number of absorbed photons per atom was $\overline{\mathcal{N}}_\mathrm{abs} \approx183$. (inset) Plot showing the effect of $\alpha^*$ for a specific ROI with $\langle od \rangle \approx 2$ (black dots: optimal $\alpha^*$, open circles: $\alpha^* \pm 0.3$ and $\pm0.6$).  The optimal $\alpha^*$ minimizes the dependence of $ \langle od \rangle$ on probe intensity~\cite{Reinaudi2007}. 
        }
	\label{fig:LBBreakdown}
\end{figure}

Figure~\ref{fig:LBBreakdown} shows such linear fits to the averages in Eq.~\eqref{eq:LambertBeerLinear} for each ROI, where the $\alpha^*_\mathrm{k}$-coefficients are given by the slopes, and the offsets provide the average optical density $od_\mathrm{k}$ in each ROI. This method has the advantage of providing uncertainties on the fitted quantities, which is used in the following. In what follows, we will refer to this as the \emph{local} calibration method.

In addition, Fig.~\ref{fig:LBBreakdown} illustrates two other methods to determine non-spatially-resolved $\alpha^*$-coefficients. The \emph{original} method~\cite{Reinaudi2007} determines the peak optical density as a function of the imaging intensity for various values of $\alpha^*$. An incorrect $\alpha^*$ leads to a clear dependence on the intensity, while the optimal value minimizes the variation as shown in the inset of Fig.~\ref{fig:LBBreakdown}. Note that Ref.~\cite{Reinaudi2007} kept the total number of photons in the imaging pulse constant, while the total number of absorbed photons is kept constant in this work. 
Moreover, a \emph{global} $\alpha^*$-coefficient can be determined by taking the averages in Eq.~\eqref{eq:LambertBeerLinear} over the entire image and performing the linear fitting technique (see orange line in Fig.~\ref{fig:LBBreakdown}). This method provides a simpler approach that focuses on calibrating the total atom number instead of the peak $od$ as in the \textit{original} method.

\subsection{Density dependence}
\label{sec:DensityCalib}

\begin{figure}[tbp]
 	\centering
 	\includegraphics[width=1\linewidth]{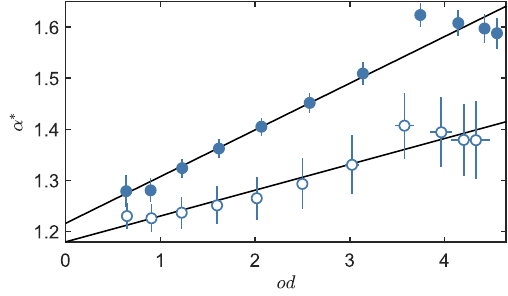}
        \caption{Dependence of the $\alpha^*$-coefficient on optical density. The solid and open points correspond to data taken with an average number of $\overline{\mathcal{N}}_\mathrm{abs}\approx183$ and $71$ absorbed photons per atom, respectively. The errors represent the uncertainty of the linear fits in Fig.~\ref{fig:LBBreakdown} without weights. The larger errors for open points show that fewer absorbed photons lead to a larger discrepancy from a linear fit (see main text). We exclude ROIs with $\langle od \rangle < 0.5$ due to the large uncertainties of these points. The lines are fits according to Eq.~\ref{eq:AlphaLinear}.
}
 	\label{fig:spatialDependency}
\end{figure}

A close inspection of Fig.~\ref{fig:LBBreakdown} shows that the magnitude of the slope increases for ROIs with larger $\langle od \rangle$, i.e., the lines with higher $y$-intercept, \emph{cf.} Eq.~\eqref{eq:LambertBeerLinear}. Moreover, the data points deviate from a linear dependence for the highest optical densities. While our analysis captures the general density dependence, a microscopic description of the imaging process is required to explain the nonlinear behavior~\cite{Veyron2022b}.

Figure~\ref{fig:spatialDependency} shows the $\alpha^*$-coefficient extracted within each ROI as a function of the corresponding optical density for two average numbers of scattered photons per atom. The $\alpha^*$ values show a clear increase as a function of optical density, and this dependence is enhanced for a larger average number of scattered photons. We attribute this to the effects outlined in Sec.~\ref{sec:Introduction}, dominated by the re-absorption of photons.

The density dependence in Fig.~\ref{fig:spatialDependency} is modelled with a linear relation 
\begin{equation}
    \label{eq:AlphaLinear}
    \alpha^* = \alpha_0 + \alpha' od.
\end{equation}
Figure~\ref{fig:spatialDependency} includes two such linear fits, which allow us to determine $\alpha^*$ for all optical densities in a given atom cloud. Thus, this realizes a spatially dependent calibration that can be applied in subsequent experiments.

To investigate the origin of the density dependence, we have simulated the effects of a non-square pulse shape, motion due to gravity during exposure, and a time-dependent detuning that depends on the number of scattered photons per atom in each voxel by numerically integrating Eq.~\eqref{eq:ScatteringRate} on a 3D grid using a finite difference approach. These simple simulations, which do \emph{not} include many-body effects such as re-absorption processes, do not lead to appreciable deviations from the expected $od$-independent $\alpha^*$-coefficient. Therefore, we conclude that many body scattering processes must be included for an accurate representation of the dynamics involved. This is corroborated by Ref.~\cite{Veyron2022}, which explores, theoretically and experimentally, the effect of incoherent many-body scattering and shows a similar linear dependence to the one used in Eq.~\eqref{eq:AlphaLinear}.

\subsection{Photon number dependence}
\label{sec:PhotonCalib}

The differences in the $\alpha^*$-coefficients obtained at various numbers of absorbed photons per atom motivate further investigation. The calibration was, therefore, repeated multiple times for different average absorbed photon numbers between $\overline{\mathcal{N}}_\mathrm{abs} \approx 38$ and $185$.

\begin{figure}[tb]
	\centering
	\includegraphics[width=1\linewidth]{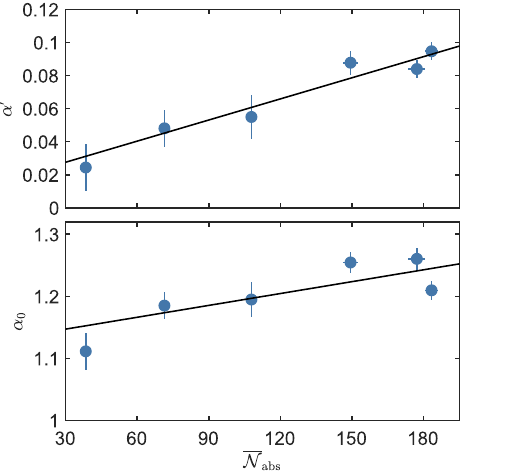}
	\caption{Gradient $\alpha'$ (top panel) and offset value $\alpha_0$ (bottom panel) of the $\alpha^*$-coefficient as a function of the average absorbed number of photons per atom $\overline{\mathcal{N}}_\mathrm{abs}$. The points show the values extracted from weighted linear fits according to Eq.~\eqref{eq:AlphaLinear} (\emph{cf.} Fig.~\ref{fig:spatialDependency}), and the error bars denote the errors from these fits. Linear fits are included in both panels.
    }
	\label{fig:Slope}
\end{figure}

Figure~\ref{fig:Slope} shows the slope $\alpha'$ and offset value $\alpha_0$ as a function of $\overline{\mathcal{N}}_\mathrm{abs}$. The slope $\alpha'$ increases to good approximation linearly for the values of $\overline{\mathcal{N}}_\mathrm{abs}$ considered here, which we attribute to the many-body scattering effects outlined in Sec.~\ref{sec:Introduction}. A similar, statistically slightly less significant linear increase is observed for $\alpha_0$. This increase as a function of the absorbed number of photons may be attributed to the resulting larger effective detuning due to the accumulated Doppler effect. Most importantly, this illustrates the importance of performing the calibration under the same conditions at which imaging in later experiments is performed (see also Sec.~\ref{sec:Discussion}). In particular, the same average number of absorbed photons should be used, and the calibration measurements should include clouds of similar average $od$.

\section{Impact of spatial calibration}

This section describes the implementation of the \emph{local} calibration and demonstrates its effect on the cloud profile, total atom number $N$, and temperature $T$. The linear dependence of $\alpha^* $ can easily be incorporated into the standard procedure for extracting the optical density. By inserting Eq.~\eqref{eq:LambertBeerSolutionWithAlpha} into Eq.~\eqref{eq:AlphaLinear} one obtains 
\begin{equation}\label{eq:directOD}
   od = \frac{\alpha_0 \ln\left(\frac{\tilde{I}_0}{\tilde{I}_\mathrm{p}}\right)+ \tilde{I}_0-\tilde{I}_\mathrm{p}}{1 - \alpha' \ln\left(\frac{\tilde{I}_0}{\tilde{I}_\mathrm{p}}\right)}
\end{equation}
Thus, one can use the calibrated values of $\alpha_0$ and $\alpha'$ to determine the $od$ directly. Using this procedure, each pixel is assigned an individual $\alpha^* $ value given by
\begin{equation}\label{eq:AlphaCorrected}
\alpha^* = \frac{\alpha_0 + \alpha'(\tilde{I}_0-\tilde{I}_\mathrm{p})}{1 - \alpha'\ln\left(\frac{\tilde{I}_0}{\tilde{I}_\mathrm{p}}\right)}.
\end{equation}

To test the effect of the \textit{local} spatial calibration, we compare it to the \textit{original} method via an analysis of clouds with a range of different atom numbers $N$ and temperatures $T$. We studied atomic clouds under 10 different conditions by gradually lowering the end-frequency of the RF evaporation and, for each end-frequency, we produced sets of $\approx 60$ identical clouds. The hottest set of clouds had $T = 1.855\pm0.006\:\mu\mathrm{K}$, $N = (3.34\pm0.02)\times10^6$ and peak $od = 4.57\pm0.02$, while the coldest set had $T = 693\pm5 \: \mathrm{nK}$, $N = (1.86 \pm 0.02)\times 10^6$, and peak $od = 6.44\pm0.07$. 
 
The clouds were imaged under conditions where the average number of absorbed photons per atom was $\overline{\mathcal{N}}_{\mathrm{abs}} =180\pm5$ and the obtained calibration coefficients were $\alpha_0=1.21$, $\alpha' = 9.5\times10^{-2}$ (\textit{local} method), $\alpha^* = 1.69$ (\textit{original} method), and $\alpha^*= 1.51$ (\textit{global} method).
 
\begin{figure}[t!bp]
	\centering
	\includegraphics[width=1\linewidth, angle=0]{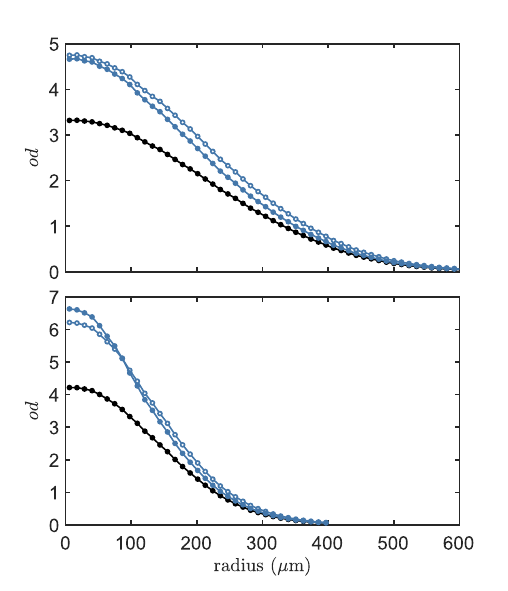}
	\caption{Comparison of the $od$ profiles for atomic clouds evaluated using three different calibration methods: the \textit{local} calibration (solid blue points), the \textit{original} calibration method used in Ref.~\cite{Reinaudi2007} (open blue circles), and no calibration ($\alpha^* = 1$, black points). The $od$ was averaged azimuthally along elliptical contours, and here, the horizontal dimension is shown. The figure shows the cloud profiles for both a hot (top, $T = 1.855\pm0.006\:\mu\mathrm{K}$) and a cold cloud (bottom,  $T = 693\pm5 \: \mathrm{nK}$).
 }
	\label{fig:DensityComparison}
\end{figure}

Figure~\ref{fig:DensityComparison} shows examples of azimuthally-averaged cloud profiles obtained with the \textit{original} method, the \textit{local} method and without calibration ($\alpha^* = 1$). The optical density was averaged along elliptical contours with a horizontal width of three pixels. The upper and lower figures show examples of clouds from the hottest and coldest set, respectively.

The \textit{local} method typically yields a lower estimated optical density in the wings of the clouds than the \textit{original} method, resulting in a narrower distribution. However, the calibrated peak $od$ can be either lower or higher depending on the temperature and actual peak $od$. To understand this behavior, it is important to note that the two calibrations were conducted on clouds with a peak $od\approx 4.5$, similar to the cloud in the upper figure. The \textit{original} method determines a single $\alpha^*$ value that minimizes the dependence of the peak $od$ dependence of the probe intensity and applies this locally optimal $\alpha^*$ on all pixels. Knowing that the appropriate $\alpha^*$ increases with $od$ it is clear from Eq.~\ref{eq:LambertBeerSolutionWithAlpha} that the \emph{original} method overestimates all optical densities below the peak $od$ used in the calibration cloud and underestimates all optical densities above.

\begin{figure}[t!]
	\centering
	\includegraphics[width=1\linewidth, angle=0]{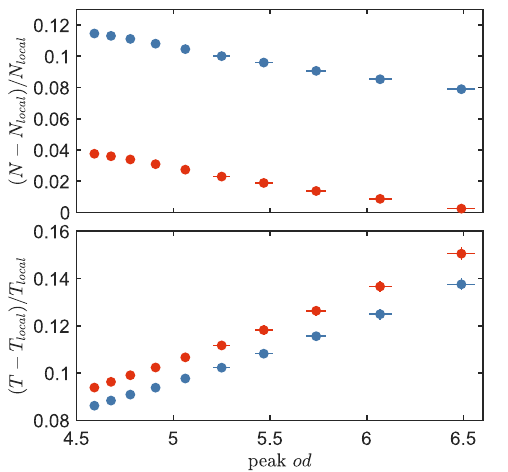}
	\caption{Relative deviations in the atom number and temperature resulting from the \textit{original} calibration~\cite{Reinaudi2007} (blue) and the \textit{global}  calibration (red) compared to the \textit{local} method ($N_\mathrm{local}$ and $T_\mathrm{local}$). The points (error bars) denote the mean values (standard deviation) from $\approx60$ clouds prepared with the same RF end-frequency. }
	\label{fig:GOvsSpatial}
\end{figure}

These effects generally lead to discrepancies between the two methods in extracting atom number, $N$, and temperature $T$. The deviations of the \textit{original} method from the \textit{local} method are shown in Fig.~\ref{fig:GOvsSpatial}. For both methods, $N$ is calculated by summing the signal in the full ROI, and $T$ is extracted from the width of a fitted Gaussian distribution. Each point is the mean value of sets of $\approx60$ clouds prepared with the same end-frequency of the RF evaporation. 

The \emph{local} method introduces corrections in the total atom number up to $\approx \!\SI{12}{\percent}$ and in temperature up to $\approx\! \SI{14}{\percent}$. For clouds with a peak $od$ close to the one used for the calibration, the \textit{original} method most dramatically overestimates the total atom number since the optical density in the wings is overestimated. When cooling further and increasing the peak $od$ a gradually increasing underestimation of the peak $od$ by the \textit{original} method begins to compensate for the overestimation in the wings. 

For temperature, however, the trend is the opposite. Here, the correction from the \textit{local} method is lowest for low peak $od$, where there is an overestimation by the \textit{original} method in the wings of the cloud. For colder clouds, this overestimation is accompanied by an underestimation of the peak $od$, leading to an increasing overestimation of the cloud width and temperature.  

Overall, the \emph{original} method is not fully accurate even for the exact cloud parameters used in the calibration, since it only performs the calibration for a specific $od$. Thus, the derived $\alpha^*$ will not necessarily render parameters like $N$ and $T$ independent of the probe intensity. The \textit{global} method suffers from similar issues, but it improves the accuracy of the total atom number since it is a calibration for the average $od$. On the other hand, the \textit{local} method addresses the density dependence of $\alpha^*$ by providing individual calibration parameters for each pixel.

\section{Practical calibration considerations}
\label{sec:Discussion} 

In order to perform an $\alpha^*$ calibration for a given experiment, one proceeds as follows: First, a set of images is taken with $\alpha^* = 1$, varying $I_\mathrm{p}$ while keeping $\overline{\mathcal{N}}_\mathrm{abs}$ constant by adjusting the imaging pulse duration. 
The data is then segmented into elliptical ROIs as shown in Fig.~\ref{fig:1ROI} and plotted as in Fig.~\ref{fig:LBBreakdown}(d). Based on this data, the density dependence of $\alpha^*$ can be obtained as a function of $od$ as shown in Fig.~\ref{fig:DensityComparison}, and the slope and offset ($\alpha'$ and $\alpha_0$) can be extracted from a linear fit. These values allow for a pixel-by-pixel determination of $\alpha^*$ from Eq.~\eqref{eq:AlphaCorrected}. The system is now calibrated, and the density profile of future experimental images can be corrected using the pixel-by-pixel correction of $\alpha^*$ in Eq.~\eqref{eq:LambertBeerSolutionWithAlpha}.

Note that the slope and intercept are dependent on the value of $\overline{\mathcal{N}}_\mathrm{abs}$ and re-calibration is necessary for different $\overline{\mathcal{N}}_\mathrm{abs}$. In practice, low $od$ values lead to large uncertainties in the fit, so only cloud ROIs with $\langle od\rangle$ above $0.5$ are used to extract $\alpha'$ and $\alpha_0$. It may be advantageous to perform the calibration for a shorter time of flight than used in the final experiment, which will enable it to cover a broader range of $od$ values. 

Moreover, it is important to reduce the effects of fringes in images that are caused by variations between the probe and reference images due to vibration of the optical elements or out-of-focus effects and astigmatism. These fringes should be minimized (e.g. using minimal separation between the probe and the reference image or using the methods in Refs.~\cite{Zhou2018, Ness2020, Jo2020}), since the local value of $\alpha^*$ in these regions will enhance the effect of the fringing. That is, the dense region in a fringe will be given a higher local $\alpha^*$, leading to an enhancement in the $od$.

While the method and data presented here profit from the atom number stabilization provided by Faraday imaging, this is not a necessary requirement to perform this type of calibration. One would find that the error bars on the data lead to a larger uncertainty in $\alpha^*$; however, the typical atom number and $od$ fluctuations in a well-functioning experiment would not preclude a spatially-resolved $\alpha^*$ calibration.

\section{Conclusion}

In conclusion, we have presented a spatial calibration technique for time-of-flight absorption images of ultracold atomic clouds. Our calibration method is \emph{local} since it assigns a calibration coefficient $\alpha^*$ for each optical density. The analysis shows a significant dependence of the calibration coefficient $\alpha^*$ on the optical density which we attribute to the many-body scattering processes and the re-absorption of photons~\cite{Veyron2022,Veyron2022b}. The calibration coefficient can readily be obtained from absorption images, and we provide a practical guide to the calibration process. Finally, it is shown that the spatial calibration technique leads to corrections in the detected atom number up to $\approx \!\SI{12}{\percent}$ and in temperature up to $\approx\! \SI{14}{\percent}$ in comparison to previous calibration techniques.

Our work highlights the aforementioned difficulties in the accurate determination of atom numbers and temperatures of dense clouds of ultracold atoms. Accurate techniques are available for small~\cite{Hume2013,Bucker2009} and mesoscopic~\cite{Muessel2013} atoms numbers. However, the best available calibration techniques for large atomic samples rely on models of light interacting with a single atom~\cite{Yu2007, Reinaudi2007}. Due to a lack of alternatives, these techniques are nonetheless commonly employed in dense atomic clouds and BECs~\cite{Veyron2022}. The technique presented here mitigates some of the difficulties by an empirical calibration strategy, and as such, it could be applied to BECs, e.g., by employing long times of flight or partial-transfer absorption imaging~\cite{Campbell2012} to achieve optical densities in the range considered here.

Our work shows the need for further theoretical developments to understand the scattering processes in ultracold dense clouds of atoms~\cite{Veyron2022b}. A more detailed understanding may explain the differences between the results reported here and those presented in Ref.~\cite{Veyron2022} for \emph{in-situ} measurements. In particular, we do not observe a universal slope of $\alpha^*$ in Eq.~\eqref{eq:AlphaLinear}, but find that this depends on the absorbed photon number. Furthermore, future experimental work should explore the applicability of our method at lower temperatures and higher densities, particularly in the BEC regime. These developments will allow for a better understanding of the many-body scattering processes that underpin atom-light interactions in dense atomic ensembles.

\begin{acknowledgments}
We acknowledge support from the Danish National Research Foundation through the Center of Excellence “CCQ”(Grantno. DNRF156), by the Novo Nordisk Foundation NERD grant (Grantno. NNF22OC0075986), and by the Independent Research Fund Denmark (Grantno. 013500205B).

\end{acknowledgments}

\bibliography{References/spatial_alpha_calibration}

\end{document}